\documentstyle[epsf,12pt]{article}
\newcommand{\yourtitle}[1]{
\mbox{}\\
\vskip 4\baselineskip
{\bf\noindent #1}\\ }
\newcommand{\youraddress}[1]{
\noindent\mbox{}\hspace{1in}\parbox[t]{4.0in}{#1}\\ }
\newcommand{\yournames}[1]{
\mbox{}\\
\mbox{}\\
\noindent\mbox{}\hspace{1in}{#1}\\ }
\newcommand{\yourabstract}[1]{
\mbox{}\\
\mbox{}\\
{\bf\noindent Abstract}\\
\begin{center}
\mbox{}\parbox[t]{5.in}{#1}
\end{center} }
\newcommand{\yoursection}[1]{
\vskip 2\baselineskip
{\bf\noindent #1}\\
\mbox{}\\
\vspace{-0.19in}}
\newcommand{\referencestyle}{
\small
\abovedisplayskip=6pt
\belowdisplayskip=6pt
\vspace{12pt}}

\begin{document}

\rightline{hep-ph/9312330}

\yourtitle{QUARK-GLUON PLASMA FREEZE-OUT \mbox{} \\
 FROM A SUPERCOOLED STATE?
\footnote{To appear in the Proceedings of the
 Workshop on Preequilibrium Parton Dynamics,
 August 1993, Lawrence Berkeley Laboratory, Berkeley, California
(ed. X. N. Wang)}\hfill\null
}
\yournames{T. Cs\"org\H o$^1$ and L. P. Csernai$^2$ \hfill\null \\ \null}
\youraddress{$^1$ KFKI RMKI \hfill\null \\
        H-1525 Budapest 114, POB 49, \hfill\null \\
        Hungary \hfill\null \\ \mbox{} \\
        $^2$ Centre for Theoretical Physics, Physics Department,\hfill\null  \\
University of Bergen,\hfill\null \\
Allegaten 55, N-5007 Bergen, Norway \hfill\null
 }
\yourabstract{
We consider time-scales of first-order deconfinement or chiral-symmetry
restoring phase transition in high energy heavy ion collisions at RHIC
and LHC energies.  Recently it was shown that the system must supercool
below $T_c$ before the nucleation of hadronic bubbles is sufficiently
rapid to overcome the expansion rate.  It is shown here that the
expected time-scales of high energy heavy ion reactions are
sufficiently short to prevent the reheating of the system to near
$T_c$.  If quark-gluon plasma is produced in these collisions, it may
have to hadronize from a  supercooled state and the hadrons produced
during rehadronization may freeze-out almost immediately.
}

\yoursection{INTRODUCTION}

Recently large experimental and theoretical programmes were launched
for studying the properties of Quantum Chromo Dynamics (QCD) at high
temperatures and energy densities \cite{QM}.  At Brookhaven National
Laboratory, reactions of gold nuclei with 100 GeV/nucleon cms energy
are expected to create a hot blob of gluons and quarks, while lead
nuclei are to be collided  at the CERN LHC with 3 TeV/nucleon energy in
the c.m. frame. According to the standard picture~\cite{bjorken}, the
colliding nuclei pass through each other  at such high energies,
leaving behind a highly excited volume filled with gluons and quarks,
which then expands mainly along the beam axis.  Numerical simulations,
based on perturbative QCD and relativistic transport theory, confirmed
this picture predicting a nearly equilibrated and baryon-free plasma of
about 150 fm$^3$ with an initial temperature of 300-350 MeV
\cite{geiger}.

The dynamics of the rehadronization of the expanding and cooling plasma
phase is very sensitive to the formation rate of hadronic bubbles
inside the plasma. The characteristic nucleation time as a function of
the temperature was found to be of the order of 100 fm/c for a
longitudinally expanding gas of gluons and massless quarks
rehadronizing into a massless pion gas, see fig. 1.  of ref.
{}~\cite{nucleation}. The input to the calculation was the calculation of
the rate for the nucleation of the hadronic phase out of the plasma
phase, which can be written as \cite{nucleation,Becker,Langer}
\begin{equation}
    I = I_0 \exp(-\Delta F_* /T),
\end{equation}
where $\Delta F_*$ is the change in the free energy of the system with
the formation of a critical size hadronic bubble, $T$ is the
temperature and $I_0$ is the prefactor.  The prefactor has recently
been calculated in a coarse-grained effective field theory
approximation to QCD~\cite{prefactor}, and was used in the calculation
of the nucleation time. The characteristic nucleation time, 100 fm/c is
rather long compared to the typical hadronic time scales of 1 fm/c.

 One has to distinguish between  the nucleation time and the  time
actually needed to  complete the transition,  \cite{actually}.  In
Fig. 2. of ref. \cite{nucleation} the temperature as a function of
(proper)time was  presented based  on the  integration of  coupled
dynamical equations  describing bubble  formation and  growth in a
longitudinally expanding Bjorken  tube.  Let's recite  the results
of these calculations which  we need for our  considerations about
the time-scales  of the  ultra-relativistic heavy  ion collisions.
If the plasma is first equilibrated at a temperature $T_0 = 2 T_c$
at time $t_0 = 3/8$ fm/c as suggested by the uncertainty principle
and by  detailed simulations  \cite{geiger}, then  the plasma will
cool according to the law $T(t) = T_0 (t_0/t)^{1/3} $ until $t_c =
8 t_0 = 3$ fm/c.  The  matter continues to cool below $T_c$  until
$T$ falls to about  0.95 $T_c$ when noticeable  nucleation begins.
When the temperature has fallen to a ``bottom'' temperature,  $T_b
= 0.8 T_c$, bubble formation and growth is sufficient to begin the
reheating the system.  This occurs at about $t_b = 7$ fm/c.   When
the temperature exceeds about 0.95 $T_c$ nucleation of new bubbles
shuts off.  The transition continues only because of the growth of
previously created bubbles.  However, the temperature must  remain
somewhat below $T_c$ in order for these bubbles to grow.  Compared
to the  idealized adiabatic  Maxwell-Boltzmann construction  which
assumes  phase  equilibrium  at  $T_c$  the finite transition rate
delays the completion of the transition by about 11 fm/c, yielding
a completion time of $t_{compl} = 50$ fm.

 Detailed calculations  including dilution  factor for  the bubble
formation,  spherical  expansion,  bubble  fusion  and varying the
values  for  the  surface  tension  do  not change the qualitative
behaviour  of  the  rehadronization  process.   According  to  the
calculations  in~\cite{zsenya},  the  time-scales  become somewhat
shorter, due  to the  fusion of  the bubbles  which increases  the
speed  of   the  transition.    Thus  bubble   fusion  brings  the
temperature versus time curve closer to the Maxwell  idealization.
A spherically expanding system cools faster than a  longitudinally
expanding  one,  and  in  both  cases  the  average  bubble radius
eventually exceeds  the radius  of the  expanding nuclear  matter,
\cite{zsenya}.   Starting  from  $T_i  =  2  T_c$ at 3/8 fm/c, the
bottoming out of the temperature is achieved by the time of  about
$t_b  =  $7  to  10  fm/c,  with  a  minimum temperature of $T_b =
0.7-0.9\   T_c$,   \cite{zsenya}.    These   numbers   are  rather
insensitive  to   whether  the   matter  expands   spherically  or
longitudinally  and  to  the  precise  numerical  values  of   the
parameters~\cite{zsenya}.

\yoursection{INDICATIONS OF SUDDEN FREEZE-OUT}

Present experiments indicate early freeze-out: i) HBT results, ii)
strange antibaryon enhancement, iii) high effective temperatures and
iv) unchanged hadronic masses.

 We  estimate  the  time-scales  available for the rehadronization
process at RHIC and LHC  using data taken at present  energies and
extrapolating them to higher energies.

 How  can  we  measure  the  freeze-out  time?   The most detailed
information available about the freeze-out surface can be obtained
by studying the sideward,  outward and longitudinal components  of
the  Bose-Einstein  correlation   function  (BECF)  at   different
rapidities and transverse momenta of the pair\cite{defs}.  We  can
measure the longitudinal radius, ${R_L}$ which is proportional  to
the  freeze-out  proper-time  $t_f$!   This  is  because  the BECF
measures  only  a  piece  of  the  longitudinally  expanding tube.
Within  this  piece   the  rapidity  distributions   belonging  to
different spatial rapidities have  to overlap, so that  pions with
similar  momenta  could  emerge.   The  size  of  this  region  is
characterized  by  $\Delta  \eta$   the  width  of  the   rapidity
distribution    at    a    fixed    value    of    the     spatial
rapidity~\cite{c73,csorgo}.   For  one  dimensional expansion, the
length of the region with  a given spatial rapidity width  is just
$t \Delta  \eta$.  The  hydrodynamical formalism  of Bose-Einstein
correlations gave the result~\cite{sinyukov}
\begin{equation}
    R_L = t_f \Delta \eta = t_f \sqrt{ T_f / m_T} , \label{e:sinyu}
\end{equation}
 where  $T_f$  is  the  freeze-out  temperature  and  $m_T$ is the
transverse   mass   of   the   pions.    The   sideward component,
$R_{T,side}$ measures the geometrical radius of the pion source at
the  freeze-out  time.   The  radius  in  the  outward  direction,
$R_{T,out}$ is generally bigger than the sideward radius since  it
is    sensitive    also    to    the    duration    of    the pion
emission~\cite{bertsch,csorgo}.  In Gaussian approximation for the
transverse distribution of pion  emission and for the  proper-time
distribution  of  the  pion  emitting  source  they  are   related
by~\cite{csorgo,pratt}
\begin{equation}
        R^2_{T,out} = R_{T,side}^2 +  \beta_T^2 \Delta \tau^2 ,
\end{equation}
 where  $\beta_T  =  (p_{T,1}  +  p_{T,2})/(E_1  +  E_2)  $ is the
transverse velocity of the pair  in LCMS \cite{defs}, and we  have
assumed that the BECF in  terms of the momentum difference  of the
pair, {\bf Q}, is parametrized in the form
\begin{equation}
        C(Q_{T,side}, Q_{T,out},Q_L)  = 1 +
        \lambda
        \exp(\, \,- \, R_{T,side}^2 Q_{T,side}^2 - R_{T,out}^2 Q_{T,out}^2
             -R_L^2 Q_L^2) \label{e:c2}
\end{equation}
 At CERN  SPS energies  (20 GeV/nucleon  in the  cms), preliminary
data of NA35 collaboration  indicate that the pion  emission might
be fully chaotic for $S+Au $ collisions ~\cite{Seyboth}.  This  is
also supported by  the RQMD simulation,  which describes both  the
NA44 and NA35 data using  a fully chaotic source, when  effects of
long-lived resonances as  well as particle  mis-identification and
detection  efficiency  cuts  are  taken  into account~\cite{RQMD}.
Thus the above parameterization of BECF-s is to be considered as a
phenomenological one  where the  intercept parameter,  $\lambda $,
takes    into    account    effects    coming    from     particle
mis-identification, acceptance cuts, long-lived resonance effects.

 According to recent NA35  preliminary data (NA35 contribution  to
the Quark Matter '93 conference) RQMD overestimates the $R_{T,S} -
R_{T,out}$ by 1. - 1.5 fm/c in the $ K_t = 200-300$ MeV/c  average
transverse momentum interval.  The duration of pion production  in
RQMD is about 3-5 fm/c, thus one may estimate the duration of pion
production  in  the  NA35  experiment  to  be about 2-4 fm/c which
includes  resonance  decay  contributions.   If  one distinguishes
between the width  of the freeze-out  times for directly  produced
pions and  resonances, and  the broadening  of the  width of  pion
emission  due  to  the  resonance  decays,  one  arrives  to   the
conclusion  that  the  duration  of  freeze-out  for  the directly
produced particles must be very short, of the order of 1 fm/c.

 Both NA35  and NA44  found that  the side,  out and  longitudinal
radii are equal  within the experimental  errors~\cite{NA35,NA44}.
This indicates that  the duration of  particle emission is  short,
$\beta_T \Delta \tau < 1$ fm, cf. eq.(3).  This is very surprising
since the resonance decays are  expected to create a larger  width
of pion  emission \cite{csorgo,weiner}.   If a  strong first order
phase transition  is present  in the  reaction, the  system has to
spend a long time  in the mixed phase  to release latent heat  and
decrease the initially high  entropy density.  This in  turn would
imply a  very large  difference between  the side  and out radius,
\cite{pratt,bertsch}.

 It  was  observed  that  the  transverse radius parameter of high
energy BECF's scales with the rapidity density as
\begin{equation}
        R_{L}=R_{T,side}=R_{T,out} = c ({dn^{\pm}\over dy})^{1/3} .
\end{equation}
 This scaling  was shown  to be  valid for  the transverse  radius
independently of  the type  and energy  of the  reaction including
UA1, AFS, E802, NA35 and NA44  data and can be related to  general
freeze-out arguments~\cite{Stock}.   The exponent  (1/3) indicates
that  pions  freeze  out  at  a  given  critical  density  and the
longitudinal radius  is proportional  to the  transverse one.  The
coefficient turned out to be 1, within NA35 and NA44 errors, as we
mentioned before.

 We  use  this  trend  in  the  data for estimating the freeze-out
proper-time  for  RHIC  and  LHC  energies.   The  proportionality
constant, $c$, was determined  to be 0.9 when  using the $C =  1 +
\lambda  \exp(-R^2  Q^2/2)$  convention  for the transverse radius
\cite{na35_91}.  Thus for our  case the proportionality factor  is
decreased by $\sqrt{2}$ which yields $c = 0.64$.

 The charged particle rapidity density is about 133 at midrapidity
for  central  $^{32}S   +  ^{238}U  $   collisions  at  CERN   SPS
corresponding to  $R_L =  4.5 \pm  0.5 $  fm implying a freeze-out
time of  $t_f =  4.5 -  6.5 fm/c$.   The charged particle rapidity
density was shown to scale with the projectile mass number in case
of symmetric collisions as
\begin{equation}
        {dn^{\pm}\over dy} =0.9 A^{\alpha} \ln({\sqrt{s}/2m_p}) ,
\end{equation}
 where the exponent $\alpha $ was  found to be in the region  $1.1
\le \alpha  \le 4/3$~\cite{satz}.   Combining these  equations the
target mass and energy  dependence of the freeze-out  time, $t_f$,
is given as
\begin{equation}
     t_f = 0.58 A^{\alpha /3}\sqrt{m_T/T_f} \ln^{1/3}({\sqrt{s}/2m_p}).
\end{equation}
 For different  high energy  heavy ion  reactions we  estimate the
freeze-out  proper-time  using   $\alpha  =  1.3$.    The  varying
transverse  mass  of  the  pions  and  the  unknown  value for the
freeze-out temperature increase  the uncertainty in  our estimate.
However, the number of pions  with a given $m_T$ is  exponentially
falling for large values of $m_T$ and then the relative number  of
pions with $m_T \ge 2 T_f$ is rather small, giving a $\sqrt{m_T/T}
\approx 1. - 1.4$.

 i) Summarizing  the above,  according to  eq.  (6),  the rapidity
density for CERN SPS lead on  lead increases by about a factor  of
3.5 when compared to the $S  + U$ reactions at the same  energies.
Extrapolating this  finding, at  RHIC gold  on gold  reactions the
rapidity density increases by a factor of about 7, at LHC lead  on
lead collisions by  a factor of  13, when compared  to the $S+  U$
reactions at  CERN SPS.  This in  turn implies  scaling factors of
1.51,  1.91  and  2.35  for  the  estimate  of the increase of the
freeze-out proper times.  As a result we obtain for the freeze-out
times 6-10 fm/c at CERN SPS  with lead on lead, 8-13 fm/c  at RHIC
gold on gold and 11-16 fm/c at LHC lead on lead collisions.

 Comparing the  time-scales necessary  to complete  the QCD  phase
transition  with  the  time-scales  obtained  from   extrapolating
present  interferometry  data  to  RHIC  and  LHC, we observe very
interesting coincidences.  If one starts with an initial state  as
suggested by the parton cascade simulations in ref.~\cite{geiger},
the critical temperature is reached by 3 fm/c after the collision.
By 10 fm/c time, which  is about the freeze-out time  according to
the  interferometry  estimate,  the  system  is  far  from   being
completely  rehadronized,   according  to   the  calculations   in
\cite{nucleation,zsenya}.  At this time, the system is still  very
close to the bottom of the temperature curve.

 ii)  In  terms  of  particle  composition,  the  idea  that   the
quark-gluon  plasma  has  to   hadronize  suddenly  in  a   deeply
supercooled state  has the  consequence that  the strange particle
composition  \cite{jozso}  and  especially  the production rate of
strange  antibaryons  as  suggested by \cite{rafelski,wa85a,wa85b}
could become a clean signature of the quark-gluon plasma formation
at  RHIC  and  LHC  energies  as  well  as at the present CERN SPS
energy.  The  WA85 collaboration  found large  production rates of
strange   antibaryons   at    CERN   SPS   sulphur    +   tungsten
interactions~\cite{wa85a,wa85b}.  The  ratio for  $\Xi^- /\Lambda$
observed by WA85 was found to be compatible with those from  other
interactions.   However,  the  ratio  $\overline  \Xi^- /\overline
\Lambda$  was  found  to  be  about  five times greater than those
obtained by the AFS collaboration, corresponding to a two standard
deviation effect.  Rafelski was able to reproduce this enhancement
only by assuming sudden rehadronization from QGP near equilibrium,
which would not change the strangeness abundance  \cite{rafelski}.
Really,  the  long  time-scale  of  the nucleation compared to the
short  time-scales  of  the  pion  freeze-out  times  at  CERN SPS
energies support  the coincidence  of the  maximal supercooling of
the QGP with freeze-out time  of 4.5-6.5 fm/c, leaving very  short
time for the strange antibaryons for reinteraction in the hadronic
gas  already  at  CERN  SPS  energies.   Spacelike detonations and
spacelike deflagrations from a supercooled baryon rich quark-gluon
plasma  were  related  to  strangeness  enhancement  at  CERN  SPS
energies in ref. ~\cite{cleymans}.

 iii)  The  latent  heat  during  such  a  sudden breakup might be
released  as  high  kinetic  energy  of  the hadrons in a timelike
deflagration~\cite{csernai_levai}.    This   is   in   qualitative
agreement with the  observation that the  multistrange antibaryons
observed by the WA85  collaboration are all at  transverse momenta
above 1.2 GeV/c, and show an effective $T_{slope} > $ 200 MeV. 
It  is known at lower  energies if one produces  a
hot fireball where resonances (deltas) are in thermal equilibrium,
after  the  freeze-out  and  the  resonance  decays  the effective
temperature for the protons (baryons) will be larger than those of
pions~\cite{harris}.  Further, the effective slope of the  baryons
will be about 10\%  lower, than the freeze-out  temperature.  Thus
we  may  expect  that  the  slope  parameters  of the multistrange
antibaryons  provide   more  information   about  the   freeze-out
temperature than  those of  the pions  (which at  the present CERN
experiments come out with more moderate slope parameters).

 iv) In dense and hot hadronic matter hadronic masses are expected
to  decrease  considerably,~\cite{hatsuda}.   Nevertheless, in the
dilepton spectra the observed masses of hadronic resonances  (e.g.
$\phi  $  )  were  identical  to  their  free  masses in heavy ion
reactions at CERN  SPS energies.  This  also can be  attributed to
simultaneous  hadronization  and  freeze-out,  where  the   medium
effects are ceased to exist, when  hadrons are formed.

 Thus from trends in interferometry data the {\it freeze-out} time
scale is short enough to  prevent reheating and the completion  of
the {\it rehadronization} of the quark-gluon plasma through bubble
formation in  the supercooled  state.  This  in turn  implies that
other  mechanisms   must  dominate   the  final   stages  of   the
hadronization.

\yoursection{DYNAMICS OF SUDDEN FREEZE-OUT}

 Using the parameters  of \cite{nucleation} the  value of the  bag
constant is $B^{1/4}= 235$ MeV, the critical temperature is  given
by  $T_c  =  169  $  MeV  and  the pressure of the supercooled QGP
vanishes  at  $T=0.98  T_c$  already.   According  to  the   above
considerations  and  ref.  \cite{zsenya}  the  temperature  of the
system in the supercooled phase reaches $T=0.7-0.9 T_c \simeq  120
- 150$ MeV.

 Observe, that at such a low temperatures the pressure of the  QGP
phase takes large negative values in the bag model.  Systems  with
negative  pressure  are  {\it  mechanically unstable}, either they
don't fill the available volume or they spontaneously cluster.

 In the  many dimensional  space of  coordinates corresponding  to
possible instabilities after crossing the borderline of  stability
on  the  phase  diagram  there  is  always one channel which opens
first.  This  usually corresponds  to spherical  configurations of
instability.  A deeper penetration into the supercooled region may
lead to the opening of other channels of instability.  These other
channels may include string-like, or cylindrical instabilities and
later layered instabilities  or spinodal decomposition.   Thus the
calculated nucleation  rate gives  an accurate  description of the
initial  hadronization  at  small  supercooling.  Furthermore, the
nucleation  rate  calculated  was  dominated  by  thermal  near  -
equilibrium  processes  and  by  the  thermal  interaction  of the
neighboring  particles,  or  thermal  damping.   This  is  a valid
assumption when  the critical  temperature is  reached, but  after
further expansion  and a  considerable supercooling,  e.g. 30\% or
more,  the  matter  is  not  so  dense any more and the collective
near-equilibrium interaction with  the surrounding matter  may not
be  the  dominant  process.   Instead quantum mechanical processes
including very few particles may dominate the transition.  For the
complete  study  of  the  reaction  dynamics and hadronization all
these aspects should be considered and the final conclusion may be
very  different  for  different  collision  energies and different
nuclei.

 The mechanical instability of the  QGP phase below 0.98 $T_c$  on
one hand and  the typical 100  fm/c nucleation times  on the other
hand are the basic reasons for the sudden rehadronization which we
propose.    It   is   understood,   that   the   expansion   in an
ultrarelativistic  heavy  ion  collision  is  so  fast,  that  the
temperature  drops  below  $T_c$  by  20-30  \%  before nucleation
becomes efficient enough to  start reheating the system.   By that
time, the QGP  phase is far  in the mechanically  unstable region.
The transition proceeds  from a mechanically  unstable phase to  a
mechanically stable and thermodynamically (meta)stable phase,  the
(superheated) hadron gas state.

 Let  us  consider  the  sudden  freeze  out  from supercooled QGP
\cite{holme,gong,aram}.   The  baryon  free  case was discussed in
\cite{gong}  in  detail  including  the  possibility of converting
latent  heat  to  final  kinetic  energy locally and instantly [in
timelike  deflagration],  while  \cite{aram}  did not include this
possibility.  Although  in ref.  \cite{gong} it  is argued  that a
superheated  hadronic  state  is  not  realizable  as  final state
[because  one  has  to  pass  the  mixed  state  on the way], this
restriction does not apply for sudden freeze-out which we consider
as a discontinuity across a hypersurface with normal  $\Lambda^\mu
\ ( \Lambda^\mu  \Lambda_\mu = +1  )$.  We can  satisfy the energy
and momentum conservation across this discontinuity expressed  via
the energy momentum tensors of the two phases, $T^{\mu\nu}$, \ \ $
( T_H^{\mu\nu} \  - T_Q^{\mu\nu} )  \Lambda_\nu \ =  \ 0 ,  $ with
entropy production in the $Q \ \rightarrow \ H $ process.

 Relativistic timelike  deflagrations are  governed~\cite{gong} by
the Taub adiabat,
\begin{equation}
{\displaystyle p_1 - p_0 \over \displaystyle X_1 - X_0 } =
{\displaystyle \omega_1 X_1 - \omega_0 X_0 \over \displaystyle X_1^2 - X_0^2},
\end{equation}
the Rayleigh-line
\begin{equation}
{\displaystyle p_1 - p_0 \over \displaystyle X_1 - X_0 } = \omega_0
\end{equation}
and the Poisson-adiabat,
\begin{equation}
{\displaystyle \sigma_1^2 \over \displaystyle \omega_1 } =
{\displaystyle R^2 \over \displaystyle X_1 }
{\displaystyle \sigma_0^2 \over \displaystyle \omega_0 },
\end{equation}
 where  $\omega_i  =  \varepsilon_i  +  p_i$  denotes the enthalpy
density and the quantity  $X_i$ is defined as  $ X_i = \omega_i  /
\omega_0  $.   The  index  $0$  refers  to the quantity before the
timelike deflagration, while $1$ refers to after deflagration.  If
we  suppose   that  the   flow  will   be  given   by  a   scaling
Bjorken-expansion before and after the timelike deflagration,  one
may  simplify  the  equations  governing the relativistic timelike
shocks,  introduced  in  ref.  ~\cite{gong}.   For  the scaling 1d
expansion the Taub adiabat reduces  to the equality of the  energy
densities on the two sides of the timelike hypersurface
\begin{equation}
        \varepsilon_1 = \varepsilon_0,
\end{equation}
 the  Rayleigh-line  becomes  an  identity and the Poisson-adiabat
simplifies to the requirement that the entropy density should not
decrease during the transition
\begin{equation}
        R = {\sigma_1 \over \sigma_0} \ge 0.
\end{equation}
 If we  start the  timelike deflagration  from a  30\% supercooled
state the  initial state  is a  mixture including  already 15-25\%
hadronic phase.  Indicating the volume fraction of hadrons in  the
initial state  by $h$,  the initial  energy density  is given by $
\varepsilon_0(T_0)    =     h    \varepsilon_H(T_0)     +    (1-h)
\varepsilon_Q(T_0)$ and the expression for the entropy  densities
is similar.  In the bag  model, the energy densities are  given as
$\varepsilon_Q = 3 \,  a_Q \, T_Q^4 +  B$, $ \varepsilon_H =  3 \,
a_H\,T_H^4$, $\sigma_Q = 4\,  a_Q \,T_Q^3$, $\sigma_H =  4\, a_H\,
T_H^3$ with coefficients $a_Q = (16 + 21 n_F / 2) \pi^2 /90 $  and
$ a_H  = 3  \pi^2 /  90$.  The  quantity $r  = a_Q/a_H $ gives the
ratio of the degrees of freedom of the phases.The temperatures  of
the initial and  final state can  be determined from  the Taub and
Poisson adiabats as
\begin{eqnarray}
T_1 & = T_H & = T_c \,\left[ {\displaystyle x -1 \over \displaystyle 3 \,
                ( 1 - (R^4 x)^{-1/3} )} \right]^{1/4}, \cr
T_0 & = T_Q & = T_c \,
                \left[ {\displaystyle x -1 \over \displaystyle 3 \,
                ( 1 - (R^4 x)^{-1/3} )} \right]^{1/4}\, (x R)^{-1/3},
\end{eqnarray}
where the ratio of the effective number of degrees of freedom is given by
\begin{equation}
        x = h + r \, (1-h).
\end{equation}
These equations provide a range for possible values of $T_H$ and $T_Q$
for a given initial hadronic fraction $h$. These bounds are given as
\begin{eqnarray}
\left[ { \displaystyle x-1 \over \displaystyle 3 } \right]^{1/4}
        & \le {\displaystyle T_H \over \displaystyle\strut T_C}  & \le
\left[ { \displaystyle x-1 \over \displaystyle 3 (1 - x^{-1/3})}
                            \right]^{1/4},\cr
    0   & \le {\displaystyle T_Q \over \displaystyle\strut T_C } & \le
\left[ { \displaystyle x-1 \over \displaystyle 3 (x^{4/3} - x)} \right]^{1/4},
\end{eqnarray}
 which  are  visualized  on  Figure  1.   Note,  that  the largest
possible values for the temperature of the hadronic phase as  well
as the minimum degree of supercooling in the initial phase-mixture
corresponds to the adiabatic $(  R = 1) $ timelike  deflagrations,
while entropy  production in  the transition  decreases both  the
final and the  initial temperatures at  a given hadronic  fraction
$h$.   If  the  transition  starts  from  a  pure quark phase, the
initial temperature  must be  at least  $ 0.7\,  T_C$ according to
Fig.  1.   However,  as  the  initial hadronic fraction approaches
unity,  the  maximum  of  the  possible  initial  temperatures for
timelike deflagrations approaches $T_C$ the critical  temperature.
Another interesting feature of Figure 1 is that for large  initial
hadronic fraction,  $h \ge  0.9$ timelike  deflagrations to  final
hadronic state with $T_H \le 0.8 \,T_C$ becomes possible.  At $T_H
= 0.8 \,T_C$ the hadronic phase is already freezed out.

          \begin{figure}
          \vglue 12pt
          \begin{center}
          \leavevmode\epsfysize=7.in
          \epsfbox{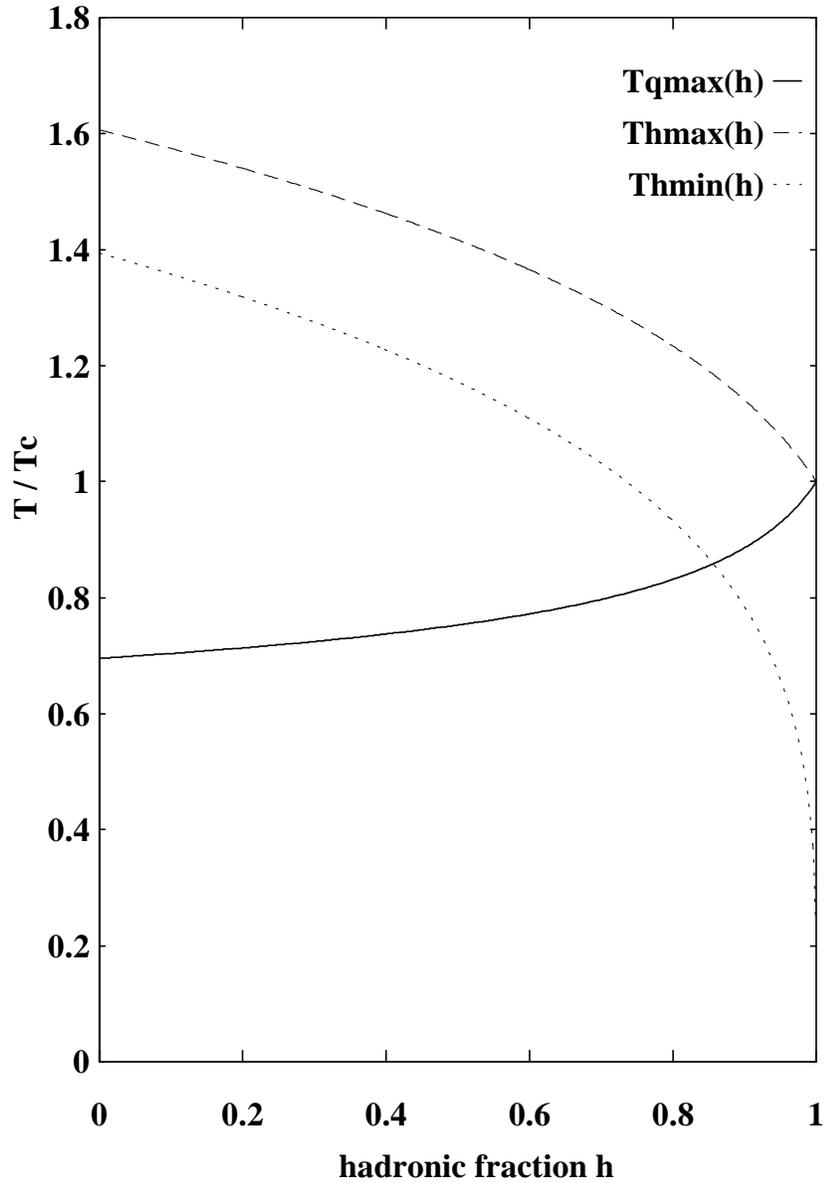}
          \end{center}
          \caption{Temperature limits for the initial and final state
           for a timelike deflagration from supercooled Quark-Gluon Plasma
          to hadron gas, $r = 37/3$. Solid line indicates the upper limit for
          the temperature of the initial QGP phase mixed with hadronic bubbles
          occupying volume fraction $h$. Dashed and dotted lines stand for
          the upper and lower limit of the temperature of the hadronic gas
          state   after the timelike deflagration, respectively.}
          \label{fig1}
          \end{figure}

          \begin{figure}
          \vglue 12pt
          \begin{center}
          \leavevmode\epsfysize=7.in
          \epsfbox{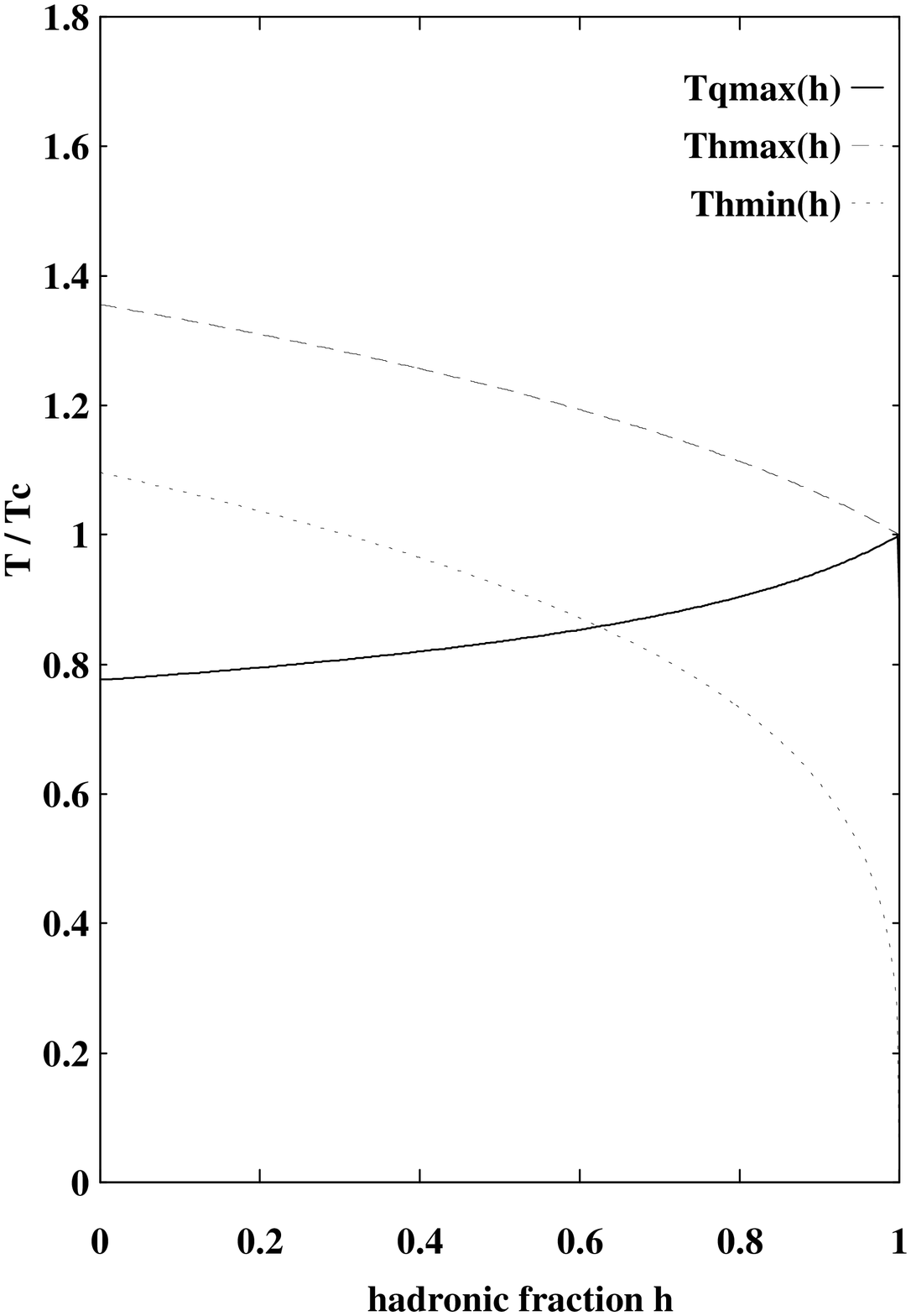}
          \end{center}
          \caption{Temperature limits for the initial and final state for
           a timelike deflagration from supercooled Gluonic Plasma to hadron
          gas, $ r = 16/3$.   Solid line indicates the upper limit for
          the temperature of the initial QGP phase mixed with hadronic bubbles
          occupying volume fraction $h$. Dashed and dotted lines stand for
          the upper and lower limit of the temperature of the hadronic gas
          state   after the timelike deflagration, respectively.}
          \label{fig2}
          \end{figure}

 Although it was pointed out that timelike deflagrations to a pion
gas at freeze-out temperature is in principle possible even in the
Bjorken model with bag equation of state, the initial temperatures
necessary for such transitions are rather low.  However, only  the
effects coming from the admixture of hadrons to the initial  state
and the  effects related  to possible  entropy production  in the
timelike deflagrations were taken  into account up to  this point.
The following mechanisms  may make such  a sudden freeze-out  more
feasible:

\begin{itemize}

 \item[-]    As    it    was    much    discussed    during   this
workshop~\cite{wang,geiger}   the   quark   degrees   of   freedom
equilibrate much slower than the gluons during the first 3 fm/c at
RHIC or LHC energies.  A hot glue scenario~\cite{shuryak} is  also
discussed  where   a  hot   gluonic  plasma   develops  from   the
preequilibrium parton collisions.  In such a plasma the number  of
degrees of freedom  is less then  in a quark-gluon  plasma, and so
the released latent heat in a timelike deflagration is also  less.
For the same initial supercooling, cooler hadronic gas states  are
reached.  This  effect is  shown in  Figure 2,  where the limiting
temperatures are shown for the final hadronic phase as well as for
the initial supercooled  phase mixture for  a pure gluonic  plasma
with $r = 16/3$.

 \item[-] Time  is needed  for the  completion of  the microscopic
quantum\-mechanical  processes  that  lead  to the sudden timelike
deflagration.   The  deflagration  front  will  have  a   timelike
thickness of about 1-2 fm/c, which leads to a further dilution  of
the matter.

 \item[-] The process of timelike deflagration may convert part of
the  latent  heat  to  the  collective kinetic energy of expansion
instead and not into the  internal thermal energy of the  hadronic
phase  \cite{csernai_levai},  leading  to  the  development  of  a
collective  transverse  flow.   The  effective  temperature of the
hadrons, as measured by the transverse momentum distribution, will
be larger than the temperature because of the transverse boost  by
the flow.  For example,  a freeze-out temperature of  $T_f \approx
140$ MeV with a transverse flow of $ \beta_T = 0.4$ results in  an
effective    temperature    $T_{eff}    \approx    \sqrt{    (1  +
\beta_T)/(1-\beta_T) } T_f = 210 $ MeV.

 \item[-] The inclusion of more realistic equation of state in the
hadronic  and  in  the  partonic  phases  might further change the
amount  of  the  latent  heat  which  together  with the transport
coefficients and the surface tension  are playing a major role  in
determining the  nucleation dynamics.   The amount  of the  latent
heat  strongly  influences   how  large  initial   supercooling  is
necessary to  reach a  given final  hadronic state.   Inclusion of
higher  hadronic  resonances  further  decreases  the ratio of the
number of degrees of freedom, $r$. \item[-] the sudden  freeze-out
will lead to a baryon excess and particularly to a strange  baryon
excess  compared  to  thermal  and  chemical  equilibrium  in  the
hadronic phase.  This reduces the number of light mesons with high
thermal velocities and thus,  advances freeze-out in the  hadronic
phase also. \end{itemize}

 Possible  mechanisms  for  such  a  non-equilibrium scenario were
considered in the combinatoric break-up model~\cite{jozso}.  Other
models like  \cite{phaser} and  \cite{cleymans} are  not providing
faster   hadronization    than   \cite{nucleation,zsenya}.     The
elaboration of the further  details of the sudden  rehadronization
of QGP in a supercooled  state is needed, especially the  study of
transverse  flow  effects  and  the  quantum processes which might
govern the timelike deflagration.

\yoursection{CONCLUSIONS}

 In summary, we considered the time-scales of rehadronization  for
a baryon-free QGP at RHIC and LHC energies.  Pion freeze out times
are estimated based on  the analysis and extrapolation  of present
high energy HBT data.  We  found that the time-scale for  reaching
the bottom of the temperature curve during the cooling process via
homogeneous nucleation \cite{nucleation} is surprisingly close  to
the time-scale of the freeze out.   We argued that the QGP has  to
complete rehadronization in a 10  - 30 \% supercooled phase  quite
suddenly, and we have shown that such a sudden process is possible
and satisfies energy and momentum conservation with  non-decreasing
entropy.

 This  rehadronization  mechanism  is  signalled  by  a  vanishing
difference  between  the  sidewards  and  outwards  components  of
Bose-Einstein  correlation  functions,  in  the observation of the
free masses of  the resonances in  the dilepton spectra,  and in a
clean strangeness signal of the QGP. Detailed calculations have to
be performed for making more quantitative predictions.

\yoursection{Acknowledgments}

 We thank J. Kapusta, D.  Boyanovsky, M. Gyulassy, P. L\'evai  and
G. Wilk for stimulating discussions.   We would like to thank  the
Organizers for invitation and support.
 This work was supported in part by the Norwegian Research Council
(NFR) under Grant  No. NFR/NAVF-422.93/008, /011,  by the NFR  and
the Hungarian Academy of Sciences exchange grant, by the Hungarian
Science Foundation  under Grants  No. OTKA-F4019  and OTKA-2973 as
well as  by the  US National  Science Foundation  under Grant  No.
PHY89-04035.


\begin{thebibliography}{99}
\referencestyle


\bibitem{QM}            Proceedings of the Quark Matter conferences,
                        especially Nucl.\ Phys.\  {\bf A498},
                        (1989), Nucl.\ Phys.\ {\bf A525}, (1991) and
                        Nucl.\ Phys.\ {\bf A544}, (1992).

\bibitem{bjorken}       J. D. Bjorken, Phys.\ Rev.\ {\bf D 27}, 140 (1983).

\bibitem{geiger}        K. Geiger and B. M\"uller, Nucl. Phys. {\bf B369},
                        600 (1992); K. Geiger,  Phys. Rev. {\bf D 46},
                        4965, 4986 (1992); ibid. {\bf 47} 133, (1993).

\bibitem{nucleation}    L. P. Csernai and J. I. Kapusta, Phys. Rev. Lett.
                        {\bf 69}, 737 (1992).

\bibitem{Becker}        R. Becker and W. D\"oring, Ann. Phys. (N.Y.) {\bf 24},
                        719 (1935).

\bibitem{Langer}        J. S. Langer, Ann. Phys. (N. Y.) {\bf 54}, 258
                        (1969).

\bibitem{prefactor}     L. P. Csernai and J. I. Kapusta, Phys. Rev.
                        {\bf D 46}, 1379 (1992).

\bibitem{actually}      J. S. Langer and A. J. Schwartz, Phys. Rev. {\bf A
                        21}, 948 (1980).

\bibitem{zsenya}        L. P. Csernai, J. I. Kapusta, Gy. Kluge and E. E.
                        Zabrodin,
                        Z. Phys. {\bf C 58}, 453 (1993).

\bibitem{HBT}           R. Hanbury-Brown and R. Q. Twiss, Nature {\bf 178},
                        1046 (1956).

\bibitem{Gyulassy}      M. Gyulassy, S. K. Kauffmann and L. W. Wilson,
                        Phys. Rev. {\bf C 20}, 2267 (1979).

\bibitem{NA35}          P. Seyboth, et al., NA35 collaboration, talk
                        presented at the meeting on {\bf Fluctuations and Soft
                        Phenomena in Multiparticle Dynamics}, Cracow, May 1993,
                        (World Scientific, Singapore, ed. R. C. Hwa, to
appear).

\bibitem{NA44}          H. B\o ggild, et al., NA44 collaboration, preprint
                        CERN-PPE-92-126, to appear in Phys.Lett. B;
                        B. L\"orstad, NA44 Collaboration, ``Recent Results
                        from the NA44 Collaboration'', in
                        Proceedings of the Budapest Workshop
                        on Relativistic Heavy Ion Collisions,
                        {\bf preprint KFKI-1993-11/A}, p. 36;

\bibitem{defs}          These components of the relative
                        momentum of the meson pair are defined in the
                        longitudinally comoving system (LCMS) belonging to
                        the pair. The longitudinal direction, $L$,  in LCMS
                        is parallel to the beam, the out direction,
                        $T,out$, in LCMS is parallel to
                        the total momentum of the pair, and the side
                        direction, $T,side$,  is transverse both
                        to the beam and the total momentum of the pair.


\bibitem{c73}
     L.P. Csernai, A.K. Holme and E.F. Staubo, in
     Proc.: {\bf ``The Nuclear Equation of State, Part B}: QCD and the
Formation
     of the Quark-Gluon Plasma'' (W. Greiner, and H. St\"ocker, eds.)
     (Plenum, 1990) p. 369-384.




\bibitem{csorgo}        T. Cs\"org\H o and S. Pratt, ``On the structure of
                        peak of Bose-Einstein Correlation Functions'',
                        Lund University preprint {\bf LU-TP-91/10} and
                        Proceedings of the Workshop on Relativistic Heavy Ion
                        Physics, KFKI Budapest preprint {\bf KFKI-1991-28/A},
                        p. 75.

\bibitem{weiner}        J. Bolz, U. Ornik, M. Plumer, B. R. Schlei,
                        R. M. Weiner, Preprint GSI-92-74,
                        Phys. Lett. {\bf B300}, 404-409 (1993)

\bibitem{pratt}         S. Pratt, Phys. Rev. {\bf D 33}, 1314 (1986).

\bibitem{sinyukov}      A. N. Makhlin and Yu. M. Sinyukov,
                        Z. Phys. {\bf C 39}, 69 (1988).

\bibitem{Seyboth}       P. Seyboth and D. Ferenc, private communications

\bibitem{RQMD}          J. P. Sullivan et al, Phys. Rev. Lett.
                        {\bf 70}, 3000 (1993).

\bibitem{bertsch}       G. Bertsch, Nucl. Phys. {\bf A498}, 173c (1989).


\bibitem{Stock}         R. Stock, Annalen der Physik {\bf 48}, 195 (1991);
                        W. A. Zajc, Nucl. Phys. {\bf A525}, 315c (1991).

\bibitem{na35_91}       P. Seyboth, NA35 collaboration
                        Nucl. Phys. {\bf A544}, 293c (1992).

\bibitem{satz}          H. Satz,
                        Nucl. Phys. {\bf A544}, 371c (1992).

\bibitem{jozso}         T. S. Bir\'o and J. Zim\'anyi,
                        Nucl. Phys. {\bf A395}, 525 (1983).

\bibitem{rafelski}      J. Rafelski, Phys. Lett. {\bf B262}, 333 (1991);
                        Nucl. Phys. {\bf A544}, 279c (1992).

\bibitem{wa85a}         S. Abatzis, et al., WA85 Collaboration,
                        Phys. Lett. {\bf B259}, 508 (1991).

\bibitem{wa85b}         S. Abatzis, et al., WA85 Collaboration,
                        Phys. Lett. {\bf B270}, 123 (1991).

\bibitem{cleymans}      N. Bili\'c, J. Cleymans, E. Suhonen and R. W. von
                        Oertzen, Phys. Lett. {\bf B311} (1993) 266;
                        N. Bili\'c, J. Cleymans, K. Reidlich and E. Suhonen,
                        preprint CERN-TH 6923/93.

\bibitem{holme}
    A. K. Holme, et al.,
    Phys. Rev.  {\bf D40}, (1989) 3735.


\bibitem{gong}
    L.P. Csernai and  M. Gong,
    Phys. Rev. {\bf D37}, (1988) 3231.


\bibitem{aram}
    J. Kapusta and A. Mekjian,
    Phys. Rev. {\bf D33}, (1986) 1304.


\bibitem{phaser}        Z. \'Arvay, T. Cs\"org\H o, J. Zim\'anyi:
                        Rehadronization of Quark-Gluon Plasma by Sequential
                        Fissioning, Proceedings of the Int.
                        Workshop on {\bf Gross Properties
                        of Nuclei and Nuclear Excitations XVIII,  Hirschegg},
                        Kleinwalsertal,
                        Austria, Jan. 15-20, 1990 (TH Darmstadt, 1990)
			p. 252;

                        Z. \'Arvay,  et al.,
                        "Production of Strange Clusters by
                        Quark-Gluon Plasma Fragmentation",
                        submitted to {\bf Zeitschrift f\"ur Physik A}.

\bibitem{cleymans}      J. Cleymans, K. Reidlich, H. Satz and E. Suhonen,
                        Z. Phys. {\bf C 58}, 347 (1993).

\bibitem{csernai_levai}
     P. L\'evai, G. Papp, E. Staubo, A.K. Holme, D. Strottman and
     L.P. Csernai, In Proc. of the Int.
     Workshop on {\bf Gross Properties
     of Nuclei and Nuclear Excitations XVIII,  Hirschegg},
     Kleinwalsertal,
     Austria, Jan. 15-20, 1990 (TH Darmstadt, 1990)
     p. 90.

\bibitem{hatsuda}       T. Hatsuda, Nucl. Phys. {\bf A544} (1992) 27c,
                        NA38 collaboration, contribution to the Quark Matter'93
                        conference (to appear in Nucl. Phys.)

\bibitem{harris}        J. Harris in "Particle Production in Highly
                        Excited Matter", H. Gutbrod and J. Rafelski eds,
                        {\bf NATO ASI series, B303} (Plenum, 1993) p. 89

\bibitem{wang}          X.-N. Wang, this proceedings;
                        K. J. Eskola and X.-N. Wang, preprint {\bf LBL-34156A},
                        nucl-th/ 9307011

\bibitem{geiger}        K. Geiger, this proceedings; K. Geiger, preprint
                        {\bf PRINT-93-0514}, contribution to the Quark
Matter'93
                        conference (to appear in Nucl. Phys.)

\bibitem{shuryak}       E. V. Shuryak,  Phys. Rev. Lett. {\bf 70},
                        2241-2244, (1993)

\end{thebibliography}
\end{document}